\documentclass[a4paper]{article}

\usepackage{INTERSPEECH2021}
\usepackage{bm}
\title{Weakly-supervised word-level pronunciation error detection in non-native English speech}
\name{Daniel Korzekwa$^1$ $^2$, Jaime Lorenzo-Trueba$^1$, Thomas Drugman$^1$, Shira Calamaro$^1$, Bozena Kostek$^2$}
\address{
  $^1$Amazon\\
  $^2$Gdansk University of Technology, Faculty of ETI, Poland}
\email{korzekwa@amazon.com}

\begin{document}

\maketitle
\begin{abstract}
We propose a weakly-supervised model for word-level mispronunciation detection in non-native (L2) English speech. To train this model, phonetically transcribed L2 speech is not required and we only need to mark mispronounced words. The lack of phonetic transcriptions for L2 speech means that the model has to learn only from a weak signal of word-level mispronunciations. Because of that and due to the limited amount of mispronounced L2 speech, the model is more likely to overfit. To limit this risk, we train it in a multi-task setup. In the first task, we estimate the probabilities of word-level mispronunciation. For the second task, we use a phoneme recognizer trained on phonetically transcribed L1 speech that is easily accessible and can be automatically annotated. Compared to state-of-the-art approaches, we improve the accuracy of detecting word-level pronunciation errors in AUC metric by 30\% on the GUT Isle Corpus of L2 Polish speakers, and by 21.5\% on the Isle Corpus of L2 German and Italian speakers.
\end{abstract}

\section{Introduction}

\begin{figure*}[!th]
  \centering
  \includegraphics[height=3.5cm]{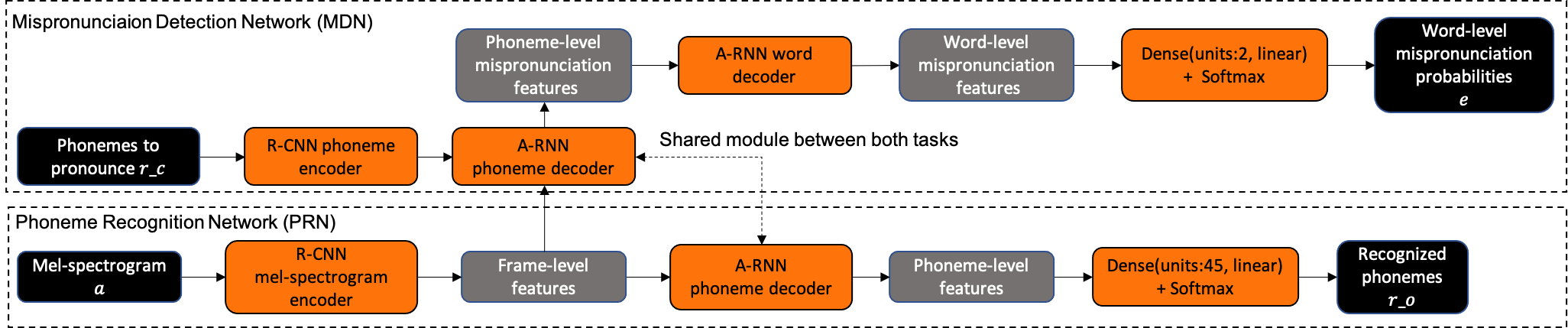}
  \caption{Neural network architecture of the WEAKLY-S model for word-level pronunciation error detection.}
  \label{fig:neural_network}
\end{figure*}

\begin{figure*}[!th]
  \centering
  \includegraphics[height=3.2cm]{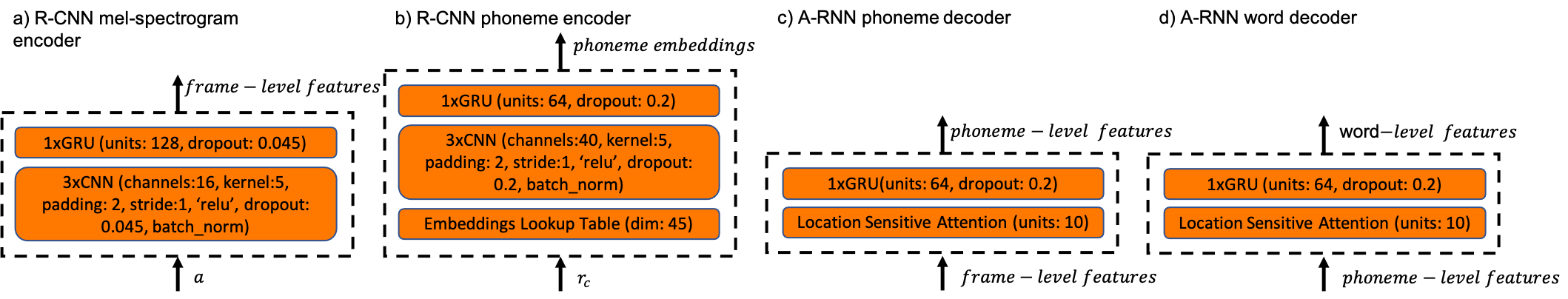}
  \caption{Details of the neural network architecture of the WEAKLY-S model for word-level pronunciation error detection.}
  \label{fig:implementation_details}
\end{figure*}

It has been shown that Computer-Assisted Pronunciation Training (CAPT) helps people practice and improve pronunciation skills \cite{neri2008effectiveness,tejedor2020assessing}. Despite significant progress over the last two decades, standard methods are still unable to detect mispronunciations with high accuracy. These methods can detect phoneme-level mispronunciations at about 60\% precision and 40\%-80\% recall \cite{leung2019cnn, korzekwa2021mispronunciation, zhang2020text}. By further raising precision we can lower the risk of providing incorrect feedback, whereas with higher recall, we can detect more mispronunciation errors.

Standard methods aim at recognizing the phonemes pronounced by a speaker and compare them with expected (canonical) pronunciation of correctly pronounced speech. Any mismatch between recognized and canonical phonemes yields a pronunciation error at the phoneme level. Phoneme recognition-based approaches rely on phonetically transcribed speech labeled by human listeners. Human-based transcription is a laborious task, especially, in the case of L2 speech where listeners have to identify mispronunciations. Sometimes, it might be even impossible to transcribe L2 speech because different languages have different phoneme sets and it is unclear which phonemes were pronounced by the speaker.

Phoneme recognition-based approaches generally fall into two categories. The first category uses forced-alignment techniques \cite{Hongyan2011, li2016mispronunciation, sudhakara2019improved, cheng2020asr} based on the work by Franco et al. \cite{franco1997automatic} and the Goodness of Pronunciation (GOP) method \cite{witt2000phone}. The GOP uses Bayesian inference to find the most likely alignment between canonical phonemes and the corresponding audio signal (forced alignment). Then, the GOP uses the likelihoods of the aligned audio signal as an indicator for mispronounced phonemes. In the second category there are methods that recognize phonemes pronounced by a speaker purely from a speech signal, and only then align them with canonical phonemes \cite{Minematsu2004PronunciationAB, harrison2009implementation, Lee2013PronunciationAV, plantinga2019towards,Sudhakara2019NoiseRG}. Techniques falling into both categories can be complemented with the use of a reference signal obtained either from a database of speech \cite{xiao2018paired, nicolao2015automatic, wang2019child} or generated from phonetic representation \cite{korzekwa2021mispronunciation, qian2010capturing}.

There are two challenges for the phoneme recognition approaches. First, phonemes pronounced by a speaker have to be recognized accurately, which has been shown to be difficult \cite{zhang2020text, chorowski2014end, chorowski2015attention, bahdanau2016end}. Second, standard approaches expect only a single canonical pronunciation of a given text, but this assumption does not always hold true due to phonetic variability of speech. In \cite{korzekwa2021mispronunciation}, we addressed these problems by modeling uncertainty in the model by incorporating a pronunciation model of L1 speech. Nonetheless, this approach still relies on phonetically transcribed L2 speech.

In this paper, we introduce a novel model (noted as WEAKLY-S) for the detection of word-level pronunciation errors that does not require phonetically transcribed L2 speech. The model produces the probabilities of mispronunciation for all words, conditioned on a spoken sentence and canonical phonemes. Mispronunciation error types include any of phoneme replacement, addition, deletion or unknown speech sound. During training, the model is weakly supervised, in the sense that we only mark mispronounced words in L2 speech and the data do not have to be phonetically transcribed. Due to the limited availability of L2 speech and the fact it is not phonetically transcribed, the model is more likely to overfit. To solve this problem, we train the model in a multi-task setup. In addition to a primary task of word-level mispronunciation detection, we use a phoneme recognizer trained on automatically transcribed L1 speech for the secondary task. Both tasks share common parts of the model, which makes the primary task less likely to overfit. Additionally, we address the overfitting problem with synthetically generated  pronunciation errors that are derived from L1 speech.

Leung et al. \cite{leung2019cnn} used a phoneme recognizer based on Connectionist Temporal Classification (CTC) for pronunciation error detection. Instead, we use an attention-based phoneme recognizer following Chorowski et al. \cite{chorowski2015attention} so that we can regularize the model by both tasks sharing a common component (attention). With a CTC-based phoneme recognizer it would not be possible because this technique does not use attention that could be shared between both tasks. Zhang et al. \cite{zhang2020text} employed a multi-task model for pronunciation assessment, but with two important differences. First, they use a Needleman-Wunsch algorithm \cite{needleman1970general} for aligning canonical and recognized sequences of phonemes, but this algorithm cannot be tuned towards sequences of phonemes. We use an attention mechanism that automatically maps the speech signal to the sequence of word-level pronunciation errors. Second, Zhang et al. detect pronunciation errors at the phoneme level and they expect L2 speech to be phonetically transcribed. This differs from our method of recognizing pronunciation errors at the word level with no need for phonetic transcriptions of L2 speech. To the best of our knowledge, this is the first approach to train word-level pronunciation error detection model that does not require phonetically transcribed L2 speech and can be optimized directly towards word-level mispronunciation detection.

\section{Proposed Model}
\label{sec:proposed_model}

\subsection{Model Definition}

The model is made of two sub-networks: \emph{i)} a word-level Mispronunciations Detection Network (MDN) detects word-level pronunciation errors $\mathbf{e}$ from the audio signal $\mathbf{a}$ and canonical phonemes $\mathbf{r_c}$, \emph{ii)} a Phoneme Recognition Network (PRN) recognizes phonemes $\mathbf{r_o}$ pronounced by a speaker from the audio signal $\mathbf{a}$ (Fig. \ref{fig:neural_network}). 

More formally, let us define the following variables: $\mathbf{a}$ - speech signal represented by a mel-spectrogram, $\mathbf{r_c}$ - canonical phonemes that the speaker was expected to pronounce, $\mathbf{r_o}$ -   phonemes pronounced, and $\mathbf{e}$ -  the probabilities of mispronouncing words in the spoken sentence. The model outputs the probabilities of word-level mispronunciation, denoted as $\mathbf{e} \sim  p(\mathbf{e}|\mathbf{a},\mathbf{r_c},\bm{\theta})$, where $\bm{\theta}$ represent parameters of the model.

We train the WEAKLY-S model in a multi-task setup. In addition to the primary task $\mathbf{e}$, we use a phoneme recognizer denoted as $\mathbf{r_o} \sim  p(\mathbf{r_o}|\mathbf{a},\bm{\theta})$ for the secondary task. The parameters $\bm{\theta}$  are shared between both tasks, which makes the MDN less likely to overfit. We define the loss function as the sum of two losses: a word-level mispronunciation loss and a phoneme recognition loss. Its formulation for the \textit{ith} training example is presented in Eq. \ref{eq:loss_function}. We train the model using two types of training data: phonetically transcribed L1 speech (both losses are used) and untranscribed L2 speech (only the mispronunciation loss is used). Having a separate loss for word-level mispronunciation lets us train the model from speech data that are not phonetically transcribed.

\begin{equation}
\mathcal{L(\bm{\theta})}=log(p(\mathbf{e}|\mathbf{a},\mathbf{r_c},\bm{\theta})) + log(p(\mathbf{r_o}|\mathbf{a},\bm{\theta}))
\label{eq:loss_function}
\end{equation}

\subsection{Neural Network Details}

\begin{figure*}[!t]
  \centering
  \includegraphics[height=2.45cm]{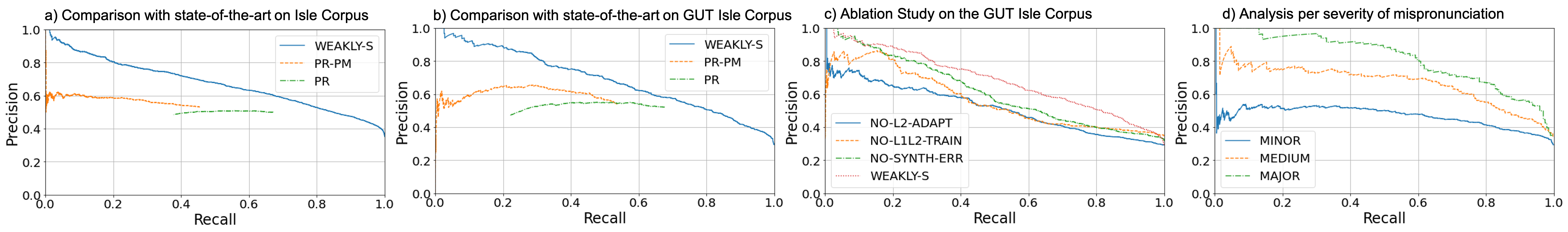}
  \caption{Precision-recall curves for the WEAKLY-S and baseline models, PR-PM and PR, (a) tested on Isle Corpus of German and Italian speakers and (b) GUT Isle Corpus of Polish speakers. (c) Ablation study on the GUT Isle corpus. (d) Analysis of mispronunciation severity levels.}
  \label{fig:precision_recall_plots}
\end{figure*}

Following Sutskever et al. \cite{sutskever2014sequence}, the MDN network encodes the mel-spectrogram $\mathbf{a}$ and the canonical phonemes  $\mathbf{r_c}$ with Recurrent Convolutional Neural Network (RCNN) encoders (Fig. \ref{fig:implementation_details}a and Fig. \ref{fig:implementation_details}b). These encoded representations are passed into an attention-based \cite{vaswani2017attention} Recurrent Neural Network (A-RNN) decoder (Fig. \ref{fig:implementation_details}c) that generates phoneme-level mispronunciation features. Phoneme-level features are transformed into word-level features (Fig. \ref{fig:implementation_details}d) based on an attention mechanism and these finally are used for computing word-level mispronunciation probabilities $\mathbf{e}$.

The PRN recognizes phonemes $\mathbf{r_o}$ pronounced by the speaker. It is similar to the attention-based phoneme recognizer by Chorowski et al. \cite{chorowski2015attention}. To generate phoneme-level features, it uses the same RCNN mel-spectrogram encoder and A-RNN decoder as the MDN. The only difference is that the A-RNN decoder is not conditioned on canonical phonemes. Phoneme-level features are transformed to the probabilities of pronounced phonemes. We added a phoneme recognition task due to the limited amount of L2 speech annotated with word-level mispronunciations. Without it, the MDN would be prone to overfitting if it was trained only on its own. By sharing common parts between both models, the PRN acts as a backbone for the MDN and makes it more robust.

The model was implemented in MxNet framework \cite{chen2015mxnet} and tuned for hyper-parameters with AutoGluon Bayesian optimization framework \cite{erickson2020autogluon}. The model was first pretrained on L1 and L2 speech corpora and then the MDN part was fine-tuned only on L2 speech data. We used the Adam optimizer with learning rate 0.001 and gradient clipping 5. Training data were segmented into buckets with batch size 32, using GluonCV \cite{guo2020gluoncv}. The A-RNN phoneme and word decoders are based on Location Sensitive Attention by Chorowski et al. \cite{chorowski2015attention}.

\section{Experiments}
\label{sec:experiments}

We present three experiments. We start with comparing our model against state-of-the-art approaches in the task of word-level mispronunciation detection. In an ablation study we analyze which elements of the model contribute the most to its performance. Finally, we analyze how the severity of pronunciation error affects the accuracy of the model.

\subsection{Speech Corpora and Metrics}

In our experiments, we use a combination of L1 and L2 English speech. L1 speech is obtained from TIMIT \cite{garofolo1993darpa} and LibriTTS \cite{Zen2019} corpora. L2 data come from the Isle \cite{atwell2003isle} corpus (German and Italian speakers) and the GUT Isle \cite{Weber2020} corpus (Polish speakers). In total, we collected 102,812 utterances, summarized in Table \ref{tab:speech_corpora}. We split the data into training and test sets, holding out 28 L2 speakers (11 German, 11 Italian, and 6 Polish) only for testing the performance of the model.

The L2 corpus of Polish speakers was annotated for word-level pronunciation errors by 5 native English speakers. Annotators marked mispronounced words and indicated their severity levels using one of the three possible values: 1 - MINOR, 2 - MEDIUM, 3 - MAJOR. The Isle corpus of German and Italian speakers comes with phoneme level mispronunciations. Words with at least one mispronounced phoneme were automatically marked as mispronounced. The Isle corpus is not mapped to severity levels of mispronunciations. In total, there are 35,555 L2 words, including 8035 mispronounced words. All data were re-sampled to 16 kHz.

We extended the train set with 292,242 utterances of L1 speech with synthetically generated pronunciation errors. We use a simple approach of perturbing phonetic transcription for the corresponding speech audio.  First, we sample these utterances with replacement from L1 corpora of human speech. Then, for each utterance, we replace phonemes with random phonemes with a probability of 0.2. In \cite{korzekwa2020detection} we found that generating incorrectly stressed speech using Text-To-Speech (TTS) improves the accuracy of detecting lexical stress errors in L2 speech. Although, as opposed to using TTS, we create pronunciation errors by perturbing the text, we expect this simpler approach should still help recognizing word-level pronunciation errors.

\begin{table}[htb]
\scriptsize
\caption{Summary of speech corpora used in experiments. * - audiobooks read by volunteers from all over the world \cite{Zen2019} }
  \label{tab:speech_corpora}
  \centering
  \begin{tabular}{lll}
    \toprule  
   Native Language & Hours & Speakers \\
    \midrule
    English & 90.47 & 640\\ 
    Unknown* & 19.91 & 285\\  
    German and Italian & 13.41 & 46\\  
    Polish & 1.49 & 12\\ 
    \bottomrule
  \end{tabular}
\end{table}

To evaluate our model, we use three standard metrics: Area Under Curve (AUC), precision and recall. The AUC metric provides an overall performance of the model accounting for all possible trade offs between precision and recall. Precision-recall plots illustrate relations between both metrics. Complementary, to analyze precision, in all our experiments we consistently fix recall at the value of 0.4 to be comparable with two baseline models that do not cover the whole range of recall values (see Section \ref{sec:sota}).

\subsection{Comparison with State-of-the-Art}\label{sec:sota}

We compare our proposed WEAKLY-S model against two state-of-the-art baselines. The phoneme recognizer (PR) model by Leung et al. \cite{leung2019cnn} is our first baseline. The PR is based on CTC loss \cite{graves2012connectionist} and it outperforms multiple alternative approaches for pronunciation assessment. The original CTC-based model uses a hard likelihood threshold applied to recognized phonemes. To compare it with two other models, following our work in \cite{korzekwa2021mispronunciation}, we replaced hard likelihood threshold with a soft threshold. The second baseline is the PR extended by a pronunciation model (PR-PM model \cite{korzekwa2021mispronunciation}). The pronunciation model accounts for phonetic variability of speech produced by native speakers, which results in higher precision of detecting pronunciation errors.

The results are presented in Fig. \ref{fig:precision_recall_plots}a, Fig. \ref{fig:precision_recall_plots}b and Table \ref{tab:accuracy_metrics}. The WEAKLY-S model turns out to outperform the second best model in AUC by 30\% from 52.8 to 68.63 and in precision by 23\% from 61.21 to 75.25 on the GUT Isle Corpus of Polish speakers. We observe similar improvements on the Isle Corpus of German and Italian speakers.

\begin{table}[!h]
%\footnotesize
\scriptsize
  \caption{Accuracy metrics of detecting word-level pronunciation errors. WEAKLY-S vs baseline models.}
  \label{tab:accuracy_metrics}
  \centering
   \begin{tabular}{llll}
    \toprule
     Model & AUC [\%] & Precision [\%,95\%CI] & Recall [\%,95\%CI] \\
    \midrule
    \multicolumn{4}{c}{\textbf{Isle corpus (German and Italian)}} \\
    PR & 55.52 & 49.39 (47.59-51.19) & 40.20 (38.62-41.81)\\ 
    PR-PM & 48.00 & 54.20 (52.32-56.08) & 40.20 (38.62-41.81)\\  
    WEAKLY-S & \textbf{67.47} & 71.94 (69.96, 73.87) & 40.14 (38.56, 41.75) \\  
	
	 \multicolumn{4}{c}{\textbf{GUT Isle corpus (Polish)}} \\
     PR & 52.8 & 54.91 (50.53-59.24) & 40.29 (36.66-44.02)\\ 
     PR-PM & 50.50 & 61.21 (56.63-65.65) & 40.15 (36.51-43.87)\\  
     WEAKLY-S & \textbf{68.63} & 75.25 (71.67-78.59) & 40.38 (37.52-43.29)\\    
    \bottomrule
  \end{tabular}
\end{table}

One difference between our model and the two baselines is that they both use the Needleman-Wunsch algorithm \cite{needleman1970general} for aligning canonical and recognized sequences of phonemes. This is a dynamic programming-based algorithm for comparing biological sequences and cannot be optimized for mispronunciation errors. Our model automatically finds the mapping between regions in the speech signal and the corresponding canonical phonemes, and then identifies word-level mispronunciation errors. In this way, we eliminate the Needleman-Wunsch algorithm as a possible source of error. 

The second difference is the use of phonetic transcriptions for L2 speech. Both baselines use automatic transcriptions provided by an Amazon-proprietary grapheme-to-phoneme model. In \cite{korzekwa2021mispronunciation} we found that for the PR and PR-PM models it is better to use automatically transcribed L2 speech for training a phoneme recognizer than not use L2 speech at all. Note that these automatic transcriptions will include phoneme mistakes for mispronounced speech. Our model does not use transcriptions of L2 speech, and instead it is guided by the word-level pronunciation errors of L2 speech in a weakly-supervised fashion.

\subsection{Ablation Study}\label{subsec:ablation_study}

We now investigate which elements of our new model contribute the most to its performance. Along with the WEAKLY-S model, we trained three additional variants, each with a certain feature removed. The NO-L2-ADAPT variant does not fine-tune the model on L2 speech, though it is still exposed to L2 speech while it is trained on a combined corpus of L1 and L2 speech. The NO-L1L2-TRAIN model is not trained on L1/L2 speech, and fine-tuning on L2 speech starts from scratch. It means that the model will not use a large amount of phonetically transcribed L1 speech data and ultimately the secondary task of the phoneme recognizer will not be used. In the NO-SYNTH-ERR model, we exclude synthetic samples of mispronounced L1 speech. It significantly reduces the amount of incorrectly pronounced words used during training from 1,129,839 to only 5,273 L2 words.

L2 Fine-tuning (NO-L2-ADAPT) is the most important factor that contributes to the performance of the model (Fig. \ref{fig:precision_recall_plots}c and Table \ref{tab:ablation_study}), with an AUC of 51.72\% compared to 68.63\% for the full model. Training the model on both L2 and L1 speech together is not sufficient. We think it is because L2 speech accounts for less than 1\% of the training data and the model naturally leans towards L1 speech. The second most important feature is training the model on a combined set of L1 and L2 speech (NO-L1L2-TRAIN), with AUC of 56.46\%. L1 speech accounts for more than 99\% of the training data. These data are also phonetically transcribed, and therefore can be used for the phoneme recognition task. The phoneme recognition task acts as a 'backbone' and reduces the effect of overfitting in the main task of detecting word pronunciation errors. Finally, excluding synthetically generated pronunciation errors (NO-SYNTH-ERR) reduces the AUC from 68.63\% to 61.54\%.

\begin{table}[th]
%\footnotesize
\scriptsize
  \caption{Ablation study for the GUT Isle corpus.}
  \label{tab:ablation_study}
  \centering
   \begin{tabular}{llll}
    \toprule
     Model & AUC [\%] & Precision [\%] & Recall [\%] \\
    \midrule
     NO-L2-ADAPT & 51.72 & 57.89 & 40.11 \\ 
     NO-L1L2-TRAIN & 56.46 & 59.73 & 40.20 \\  
     NO-SYNTH-ERR & 61.54 & 67.22 & 40.38 \\  
     WEAKLY-S & \textbf{68.63} & 75.25  & 40.38 \\    
    \bottomrule
  \end{tabular}
\end{table}

\subsection{Severity of Mispronunciation}\label{sec:severity}

When providing feedback to the L2 speaker about mispronounced words, we want to reflect the severity of mispronunciation, in order to focus on more severe errors and not report them all at once. We segment pronunciation errors into three categories: LOW, MEDIUM and HIGH, based on an inter-tester agreement of annotating sentences for word-level mispronunciations. Mispronounced words with less than 40\% inter-tester agreement belong to the LOW category, between 40\% and 80\% to MIDDLE, and over 80\% to HIGH. We validated that the proposed inter-tester agreement bands are well correlated with explicit listener opinions on the severity of mispronunciation, as shown in Table \ref{tab:inter-tester-agreement-by-severity}. This result shows that data on mispronunciation severity can be derived automatically, without the need to collect it.

\begin{table}[th]
%\footnotesize
\scriptsize
  \caption{Severity of mispronunciation by inter-tester agreement for the GUT Isle Corpus. 1 - MINOR, 2 - MEDIUM, 3 - MAJOR.}
  \label{tab:inter-tester-agreement-by-severity}
  \centering
   \begin{tabular}{ll}
    \toprule
     Inter-tester agreement & Severity [mean and 95\% CI ]  \\
    \midrule
     LOW (Less than 40\%) & 1.32 (1.28-1.35)  \\ 
     MEDIUM (Between 40\% and 80\%) & 1.58 (1.54-1.62)  \\  
     HIGH( Higher than 80\%) & 2.08 (2.03-2.13)  \\    
    \bottomrule
  \end{tabular}
\end{table}

We aim at detecting the words of HIGH inter-tester agreement with higher precision to provide more relevant feedback to L2 speakers. To make AUC, precision, and recall metrics comparable between different levels of inter-tester agreement, we enforce the ratio of mispronounced words across all categories to the same level of 29.2\% by randomly down-sampling correctly pronounced words. This value is the proportion of mispronounced words across all inter-tester agreement levels in the GUT Isle Corpus. We observe that we can detect pronunciation errors of HIGH inter-tester agreement with 91.67\% precision at 40.38\% recall (Fig. \ref{fig:precision_recall_plots}d and Table \ref{tab:severity_of_mispronunciation}). By segmenting pronunciation errors into three difference bands, we can report to a language learner only the errors of HIGH inter-tester agreement, and improve their learning experience.

\begin{table}[th]
%\footnotesize
\scriptsize
  \caption{Accuracy metrics for different severity levels of mispronunciation for the GUT Isle Corpus.}
  \label{tab:severity_of_mispronunciation}
  \centering
   \begin{tabular}{llll}
    \toprule
     Inter-test agreement & AUC [\%] & Precision [\%] & Recall [\%] \\
    \midrule
     LOW & 46.99 & 51.84 & 40.48 \\ 
     MEDIUM & 66.90 & 71.89 & 40.80 \\  
     HIGH & 81.48 & 91.67  & 40.31 \\    
    \bottomrule
  \end{tabular}
\end{table}

\section{Conclusions and Future Work}
\label{sec:conclusions}

We proposed a model for detecting pronunciation errors in English that can be trained from L2 speech labeled only for word-level mispronunciations. The data do not have to be phonetically transcribed. The model outperforms state-of-the-art models in AUC metric on the GUT Isle Corpus of Polish speakers and the Isle Corpus of German and Italian speakers. The limited amount of L2 speech and the lack of phonetically transcribed speech makes this model prone to overfitting. We overcame this issue by proposing a multi-task training with two tasks: a word-level pronunciation error detector trained on L1 and L2 speech, and a phoneme recognizer trained on L1  speech. The most important factors that contribute to the model accuracy are:  \emph{i)} fine-tuning on L2 speech,  \emph{ii)} pre-training on a joined corpus of L1 and L2 speech, and  \emph{iii)} use of synthetically generated pronunciation errors.  

The level of inter-tester agreement in annotating pronunciation errors correlates with explicit human opinions about the severity of mispronunciation. By detecting pronunciation errors only for high inter-tester agreement, we may significantly lower the number of false positives reported to a language learner. 

In the future, we will experiment with discrete representation of the latent phoneme space such as Vector-Quantized Variational-Auto-Encoder (VQ-VAE) \cite{chorowski2019unsupervised, van2017neural}, which should fit better to discrete nature of phonemes. We plan to generate synthetic mispronounced speech, which is motivated by our recent work on using speech synthesis for generating speech errors in the related task of lexical stress error detection \cite{korzekwa2020detection}.

\bibliographystyle{IEEEtran}

\bibliography{mybib}

% \begin{thebibliography}{9}
% \bibitem[1]{Davis80-COP}
%   S.\ B.\ Davis and P.\ Mermelstein,
%   ``Comparison of parametric representation for monosyllabic word recognition in continuously spoken sentences,''
%   \textit{IEEE Transactions on Acoustics, Speech and Signal Processing}, vol.~28, no.~4, pp.~357--366, 1980.
% \bibitem[2]{Rabiner89-ATO}
%   L.\ R.\ Rabiner,
%   ``A tutorial on hidden Markov models and selected applications in speech recognition,''
%   \textit{Proceedings of the IEEE}, vol.~77, no.~2, pp.~257-286, 1989.
% \bibitem[3]{Hastie09-TEO}
%   T.\ Hastie, R.\ Tibshirani, and J.\ Friedman,
%   \textit{The Elements of Statistical Learning -- Data Mining, Inference, and Prediction}.
%   New York: Springer, 2009.
% \bibitem[4]{YourName17-XXX}
%   F.\ Lastname1, F.\ Lastname2, and F.\ Lastname3,
%   ``Title of your INTERSPEECH 2021 publication,''
%   in \textit{Interspeech 2021 -- 20\textsuperscript{th} Annual Conference of the International Speech Communication Association, September 15-19, Graz, Austria, Proceedings, Proceedings}, 2020, pp.~100--104.
% \end{thebibliography}

\end{document}